\documentclass[12pt]{iopart}
\usepackage{iopams}  
\usepackage{graphicx}
\usepackage{color}
\usepackage{cite}
\eqnobysec

\def\keywords#1{\vspace{10pt}
     \begin{indented}
     \item[]\rm Keywords: #1\par
     \end{indented}}

\def\be{\begin{equation}}
\def\ee{\end{equation}}
\def\bea{\begin{eqnarray}}
\def\eea{\end{eqnarray}}
\def\CF{{\mathcal F}}
\def\CG{{\mathcal G}}
\def\CH{{\mathcal H}}

\def\CZ{{\mathcal Z}}
\def\kBT{{k_{\mathrm B}T}}
\def\ts{\tilde{s}}
\def\tK{\widetilde{K}}
\def\tH{\widetilde{H}}
\def\tY{\widetilde{Y}}
\newcommand{\tens}[1]{\boldsymbol{\mathsf{#1}}}

\begin{document}
\jl{1}

\title[Potts model with multi-site interactions]{One-dimensional \boldmath{$q$}-state Potts model 
with multi-site interactions}

\author{Lo\"\i c Turban}

\address{Groupe de Physique Statistique, Institut Jean Lamour, Universit\'e de Lorraine, CNRS (UMR 7198), 
Vand\oe uvre l\`es Nancy Cedex, F-54506, France} 

\ead{loic.turban@univ-lorraine.fr}

\begin{abstract} 
A one-dimensional (1D) $q$-state Potts model with $N$ sites, $m$-site interaction $K$ in a field $H$ 
is studied for arbitrary values of $m$. Exact results for the partition function and the two-point correlation 
function are obtained at $H=0$. The system in a field is shown to be self-dual. Using a change of Potts 
variables, it is mapped onto a standard 2D Potts model, with first-neighbour interactions $K$ and $H$, 
on a cylinder with helical boundary conditions (BC). The 2D system has a length $N/m$ and a transverse size $m$. 
Thus the Potts chain with multi-site interactions is expected to develop a 2D critical singularity along 
the self-duality line, $(\e^{qK}-1)(\e^{qH}-1)=q$, when $N/m\to\infty$ and $m\to\infty$.
\end{abstract}


\submitto{J. Phys. A: Math. Theor.}

\section{Introduction}

The standard Potts model is a lattice statistical model with pair interactions between $q$-state 
variables attached to neighbouring sites~\cite{potts52,wu82a}. 
Multi-site Potts models can be constructed by extending to an arbitrary number of states existing
multispin Ising models for which $q=2$. In this way, a self-dual three-site Potts model 
on the triangular lattice was introduced by Enting~\cite{enting75,deng10}, which corresponds to the 
Baxter-Wu model~\cite{baxter73a,baxter74} when $q=2$. Similarly, a 2D self-dual Potts model with 
$m$-site interactions in one direction and $n$-site interactions in the 
other~\cite{turban82a,alcaraz86,zhang93} follows from the Ising version with $n=1$~\cite{turban82b}. 

Multi-site interactions may be generated from two-site interactions in a position-space renormalisation 
group transformation and thus have to be included in the initial Hamiltonian. In this way Schick 
and Griffiths have introduced a three-state Potts model on the triangular lattice with two- 
and three-site interactions~\cite{schick77}. For any value of $q$ it can been reformulated as 
a standard $q$-state Potts model with two-site interactions on a 3-12 lattice~\cite{wu82b}. 
When the three-site interactions are restricted to up-pointing triangles, the model is 
self-dual~\cite{baxter78,enting78} and related to a 20-vertex model~\cite{baxter78,wu80}. 

Extending the results of Fortuin and Kasteleyn~\cite{fortuin72} for pair interactions, 
a random-cluster representation for Potts models with multi-site interactions has been 
introduced~\cite{grimmett94,chayes97,chayes98} and exploited in Monte Carlo simulations~\cite{matty15}.

Multi-site interactions enter naturally when the site percolation process is formulated as a Potts model 
in the limit $q\to1$~\cite{giri77,kunz78,essam79,kasai80,temperley82}. Various Potts multi-site 
interactions have also been used to model conformational transitions in polypeptide 
chains~\cite{goldstein84,ananikyan90,schreck10,badasyan13}.

With $s_j=0,1,\ldots,q-1$ denoting a $q$-state Potts
variable attached to site $j$, a multi-site interaction can take one of the following forms
\be
(a)\quad -K\prod_{j=1}^{m-1}\delta_{s_j,s_{j+1}}\,,\qquad
(b)\quad -K\delta_q\left(\sum_{l=0}^{m-1}s_{j+l}\right)\,,
\label{ab}
\ee
where $\delta_{n,n'}$ is the standard Kronecker delta and $\delta_q(n)$ is a Kronecker delta modulo $q$. 
When $K>0$ the ground state is $q$-times degenerate in the first case (the standard one) 
whereas the degeneracy depends on $m$ and is given by $q^{m-1}$ in the second case.
As an example, when $q=m=3$ the degenerate ground states are the following ones: 
\be
(a)\quad\left\{
\begin{array}{l}
000\\
111\\
222
\end{array}
\right.
\,,\qquad
(b)\quad\left\{
\begin{array}{lll}
000&012&210\\
111&120&021\\
222&201&102
\end{array}
\right.
\label{dgs}
\ee

In the present work we generalize for $q$-state Potts variables some results 
recently obtained for the 1D Ising model with multispin interactions~\cite{turban16}.
The Hamiltonian of the $q$-state Potts chain takes the following form:
\be\fl
-\beta\CH_N[\{s\}]=K\sum_j \left[q\delta_q\left(\sum_{l=0}^{m-1}s_{j+l}\right)\!-\!1\right]
+H\sum_j\left[q\delta_q\left(s_j\right)\!-\!1\right]\,,
\quad \beta=(\kBT)^{-1}.
\label{Hs}
\ee
We assume ferromagnetic interactions $K\geq0$ and $H\geq0$, too. The Kronecker delta modulo $q$ is given by:
\be
\delta_q(s)=\frac{1}{q}\sum_{k=0}^{q-1}\exp\left(\frac{2i\pi ks}{q}\right)=\left\{
\begin{array}{ll}
1 & \mbox{when } s=0 \pmod{q}\\
\ms
0 & \mbox{otherwise}
\end{array}
\right.\,.
\label{delq}
\ee
Introducing the Potts spins~\cite{mittag71,solyom81}
\be 
\sigma_j=\exp\left(\frac{2i\pi s_j}{q}\right)\,,
\label{sigmaj}
\ee
the Hamiltonian in~\eref{Hs} can be rewritten 
as~\footnote[1]{One may also express the Potts interaction using clock angular variables (see appendix A).}:
\be
-\beta\CH_N[\{\sigma\}]=K\sum_j\sum_{k=1}^{q-1}\prod_{l=0}^{m-1}\sigma_{j+l}^k+H\sum_j\sum_{k=1}^{q-1}\sigma_j^k\,.
\label{Hsigma}
\ee
When $q=2$, $\sigma_j=\pm1$, $k=1$ and the Ising multispin Hamiltonian studied in~\cite{turban16} is recovered, 
which {\it a posteriori} justifies the choice of interaction $(b)$ in~\eref{Hs}.

The zero-field partition function of the Potts chain 
with $m$-site interaction $K$ is obtained for free BC in section~2 and for periodic BC in section~3. 
The periodic BC result allows a determination of the eigenvalues of $\tens{T}^m$ where $\tens{T}$ 
is the site-to-site transfer-matrix. The two-site correlation function is calculated in section~4. 
In section~5 the system with periodic BC is shown to be self-dual when the external field $H$ is turned on. 
In section~6 the system with free BC is mapped onto a standard 2D Potts model with first-neighbour 
interactions $K$ and $H$, length $N/m$ and transverse size $m$. The mapping of 1D Potts models with $m$-site 
and $n$-site interactions is discussed in section~7. The conclusion in section~8 is followed by 4 appendices.

\section{Zero-field partition function for free BC}
\begin{figure}[t!]
\begin{center}
\includegraphics[width=9.3913cm,angle=0]{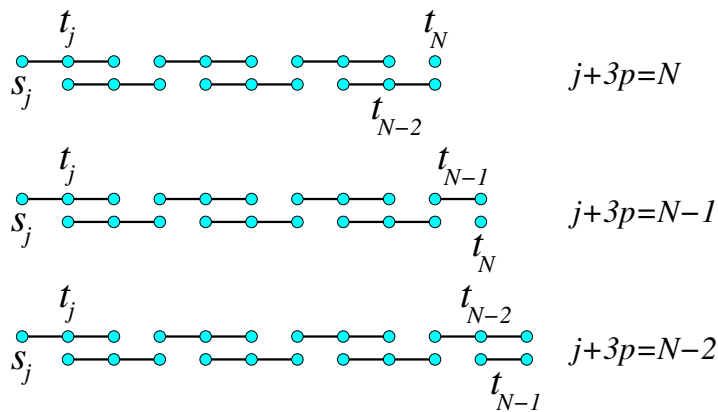}
\end{center}
\vglue -.3cm
\caption{$t$-variables entering into the expression~\eref{sj} of $s_j$ for $m=3$ and 
different values of the distance $N-j$ from the end of the chain. The $t$-variables, defined in~\eref{tj}, are the sums 
of $m$ $s$-variables (circles) with the convention $s_i=0$ when $i>N$. 
The second lines are subtracted from the first so that only $s_j$ is remaining.
\label{fig1}  
}
\end{figure}

With free BC the zero-field Hamiltonian of a chain with $N$ Potts spins, with $m$-site interaction $K$, takes the following form  
\be
-\beta\CH_N^{(f)}[\{s\}]=K\sum_{j=1}^{N-m+1}\left[q\delta_q\left(\sum_{l=0}^{m-1}s_{j+l}\right)-1\right]
\label{Hsf}
\ee
when written in terms of the Potts variables $s_j$. Let us introduce the new Potts variables $t_j=0,\ldots,q-1$ defined as
\be
t_j=\sum_{l=0}^{m-1}s_{j+l}\pmod{q}\,,\qquad j=1,\ldots,N\,,
\label{tj}
\ee
with the convention $s_i=0$ when $i>N$ in \eref{tj}. Note that the relationship between old and new variables 
is one-to-one with the inverse transformation given by (see figure~\ref{fig1}):
\be\fl
s_j=\sum_{r=0}^p(t_{rm+j}-t_{rm+j+1})\pmod{q}\,,\qquad j+pm=N-l\,,\quad l=0,\ldots,m-1\,.
\label{sj}
\ee
Using \eref{tj} in \eref{Hsf} one obtains a system of $N-m+1$ non-interacting Potts spins in a field $K$ with 
\be
-\beta\CH_N^{(f)}[\{t\}]=K\sum_{j=1}^{N-m+1}[q\delta_q(t_j)-1]\,.
\label{Htf}
\ee
The canonical partition function is easily obtained and reads: 
\bea 
\CZ_N^{(f)}&=\Tr_{\{t\}}\e^{-\beta\CH_N^{(f)}[\{t\}]}=\prod_{j=1}^{N-m+1}\Tr_{t_j}\e^{K[q\delta_q(t_j)-1]}
\prod_{j=N-m+2}^N\Tr_{t_j}1\nonumber\\
&=q^{m-1}\left[\e^{(q-1)K}+(q-1)\e^{-K}\right]^{N-m+1}\,.
\label{ZNf}
\eea
Note that although only $N-m+1$ new variables enter into the expression of the transformed Hamiltonian~\eref{Htf}, one 
has to trace over the $N$ Potts variables $t_j$ in~\eref{ZNf}. When $q=2$ 
the Ising result (equation~(2.6) in~\cite{turban16}) is recovered.

The free energy can be decomposed as follows
\be
\CF_N^{(f)}=-\kBT\ln\CZ_N^{(f)}=Nf_b+\CF_s(m)\,,
\label{Ff}
\ee
where the bulk free energy per site 
\be
f_b=-\kBT\ln\left[\e^{(q-1)K}+(q-1)\e^{-K}\right]\,,
\label{fb}
\ee
does not depend on $m$ whereas the surface contribution
\be
\CF_s(m)=(m-1)\kBT\,\ln\left\{\frac{\exp[(q-1)K]+(q-1)\exp(-K)}{q}\right\}\,,
\label{Fs}
\ee
is $m$-dependent.

\section{Zero-field partition function for periodic BC}
Let us now evaluate the partition function for a periodic chain with $N$ sites and $m>1$. To simplify the discussion we consider 
only the case where $N$ is a multiple of $m$. Then the Hamiltonian takes the following form
\be
-\beta\CH_{N=pm}^{(p)}[\{s\}]=K\sum_{j=1}^{N=pm}\left[q\delta_q\left(\sum_{l=0}^{m-1}s_{j+l}\right)-1\right]\,,
\label{Hsp}
\ee
with $s_{N+j}=s_j$. Making use of the change of variables \eref{tj}, it can be rewritten as:
\be
-\beta\CH_{N=pm}^{(p)}[\{t\}]=K\sum_{j=1}^{N=pm}[q\delta_q(t_j)-1]\,.
\label{Htp}
\ee
With periodic BC the correspondence between $\{s\}$ and $\{t\}$ Potts configurations is no longer one-to-one
and the new variables have to satisfy a set of $m-1$ constraints~\cite{turban93,garrod95,turban16,mueller17}. 

There are several $\{s\}$ configurations leading to the same $\{t\}$. 
One of these configurations, $\{s'\}$, is obtained by changing $s_j$ into 
\be
{s'}_j=s_j+\Delta_j\pmod{q}\,,\qquad j=1,\ldots,N\,.
\label{s'}
\ee
where the shifts $\Delta_j=0,\ldots,q-1$ have to satisfy some constraint.
Let us first consider $t_1=\sum_{l=0}^{m-1}s_{l+1}\pmod{q}$. One can freely choose the first $m-1$ shifts 
($q^{m-1}$ choices) and $t_1$ keeps its value when $\Delta_m$ is such that $\sum_{l=0}^{m-1}\Delta_{l+1}=0\pmod{q}$. Since $t_j$ and $t_{j+1}$ have      
the shifts $\Delta_{j+l}$ ($l=1,\ldots,m-1$ in common, the value of $\Delta_j$ leaving $t_j$ invariant 
is equal to the value of $\Delta_{j+m}$ leaving $t_{j+1}$ invariant. When $N=pm$ a periodic repetition 
with period $m$ of the first $m$ shifts acting on $\{s\}$ leaves $\{t\}$ invariant. Thus there are 
$q^{m-1}$ Potts configurations $\{s'\}$ leading to 
the same $\{t\}$~\footnote[2]{Note that the initial configuration, $\{s\}$, corresponding 
to $\Delta_j=0$ $\forall j$, is taken into account.}. 
When $\{s\}$ is a ground-state configuration, $q^{m-1}$ gives the ground-state degeneracy.

In the following we shall make use of the Potts spin variables:
\be\fl 
\tau_j=\exp\left(\frac{2i\pi t_j}{q}\right)=\prod_{l=0}^{m-1}\sigma_{j+l}\,,
\qquad\tau_j^*=\exp\left(\!-\frac{2i\pi t_j}{q}\right)=\tau_j^{q-1}\,,\quad \tau_j\tau_j^*=\tau_j^q=1\,.
\label{tauj}
\ee
According to~\eref{delq} one has:
\be
\Tr_{\tau_j}\tau_j^k
=\sum_{t_j=0}^{q-1}\exp\left(\frac{2i\pi kt_j}{q}\right)=q\delta_q(k)\,.
\label{Trtau}
\ee
For later use, note that the Boltzmann factor
\be
\e^{-\beta\CH_N^{(p)}[\{t\}]}=\prod_{j=1}^{N=pm}\left[\e^{-K}+\left(\e^{(q-1)K}-\e^{-K}\right)\delta_q(t_j)\right]\,,
\label{embH}
\ee
can be rewritten as
\bea
\e^{-\beta\CH_N^{(p)}[\{\tau\}]}&=\e^{-NK}\prod_{j=1}^{N=pm}\left[1+\frac{\e^{qK}-1}{q}
\sum_{k=0}^{q-1}\tau_j^k\right]\nonumber\\
&=\left(\frac{\e^{-K}}{q}\right)^{\!N}\,\prod_{j=1}^{N=pm}\left[\e^{qK}+q-1)+\left(\e^{qK}\!-1\right)
\sum_{k=1}^{q-1}\tau_j^k\right]\,,
\label{embH1}
\eea
using~\eref{delq} and~\eref{tauj},

Let us consider the product of Potts spins
\be
P_i=\prod_{r=0}^{p-1}\tau_{rm+i}\tau^*_{rm+i+1}\,,\qquad i=1,\ldots,m-1\,.
\label{Pi}
\ee
Making use of
\be\fl
\tau_{rm+i}\tau^*_{rm+i+1}=\sigma_{rm+i}\left(\prod_{l=1}^{m-1}\sigma_{rm+i+l}\sigma^*_{rm+i+l}\right)\sigma^*_{(r+1)m+i}
=\sigma_{rm+i}\sigma^*_{(r+1)m+i}
\label{tt}
\ee
and taking into account the periodic BC, one obtains the constraints
\be
P_i=\prod_{r=0}^{p-1}\sigma_{rm+i}\sigma^*_{(r+1)m+i}=1\,,\qquad i=1,\ldots,m-1\,,
\label{Pi1}
\ee
to be satisfied by the $\tau$-configurations in~\eref{Pi}.
When $m>2$ other constraints can be constructed, for instance from $\tau_{rm+i}\tau^*_{rm+i+2}$, 
but these are automatically satisfied since they can be written as products of the fundamental 
ones: $\tau_{rm+i}\underbrace{\tau^*_{rm+i+1}\tau_{rm+i+1}}_{1}\tau^*_{rm+i+2}$.

Thus with the new Potts spin variables, taking the constraints into account, the partition function
is given by:
\be
\CZ_{N=pm}^{(p)}=q^{m-1}\Tr_{\{\tau\}}\e^{-\beta\CH_N^{(p)}[\{\tau\}]}\prod_{i=1}^{m-1}\delta_{P_i,1}\,.
\label{ZNp}
\ee
To go further we need an explicit expression for the Kronecker delta, $\delta_{P_i,1}$. Consider the geometric series
\be
f(X)=\sum_{k=0}^{q-1}X^k=\frac{1-X^q}{1-X}\,,
\label{gs}
\ee
it vanishes when $X$ is a $q$th root of unity other than 1 and is equal to $q$ when $X=1$.
Since $P_i$ in~\eref{Pi} is a $q$th root of unity, the constraint can be written as (cf.~\eref{delq})
\be
\delta_{P_i,1}=\frac{1}{q}\sum_{k=0}^{q-1}P_i^k=\frac{1}{q}\sum_{k=0}^{q-1}\prod_{r=0}^{p-1}\tau_{rm+i}^k\tau^{q-k}_{rm+i+1}\,,
\label{delPi}
\ee
where~\eref{tauj} has been used. The partition function in~\eref{ZNp} now takes the following form:
\be\fl
\CZ_{N=pm}^{(p)}=\Tr_{\{\tau\}}\e^{-\beta\CH_N^{(p)}[\{\tau\}]}
\prod_{i=1}^{m-1}\left(1+\sum_{k=1}^{q-1}P_i^k\right)\,,\qquad
P_i^k=\prod_{r=0}^{p-1}\tau_{rm+i}^k\tau^{q-k}_{rm+i+1}\,.
\label{ZNp1}
\ee
The first product has the following expansion:
\be
\fl\prod_{i=1}^{m-1}\left(1+\sum_{k=1}^{q-1}P_i^k\right)=1+\sum_{i,k}P_i^k
+\!\!\!\sum_{i<i',k,k'}\!\!\!P_i^kP_{i'}^{k'}
+\!\!\!\!\!\!\!\!\sum_{i<i'<i'',k,k',k''}\!\!\!\!\!\!\!\!P_i^kP_{i'}^{k'}P_{i''}^{k''}+\cdots+\prod_iP_i^{q-1}.
\label{expan}
\ee
The expression of $P_i^k$ in~\eref{ZNp1} is periodic with period $m$. There are two consecutive 
Potts spins contributing to the product for each period and the sum of their exponents vanishes modulo $q$. 
Besides 1 the expansion~\eref{expan} generates terms 
containing from $l=2$ to $m$ spins for each period with ${m\choose l}$ possible spatial 
configurations $\{\alpha_l\}$. These spatial configurations are labelled by the $l$ spin exponents, 
each varying from $1$ to $q-1$ with a sum which remains vanishing modulo $q$ in the products, 
due to the Potts spins properties~\eref{tauj}. 
As shown in appendix~B, for $l$ spins the number $\nu_l$ of allowed exponent distributions is given by:
\be
\nu_l=\frac{1}{q}\left[(q-1)^l+(-1)^l(q-1)\right]\,.
\label{nul}
\ee
Combining these results, the expansion can be written as
\be
\prod_{i=1}^{m-1}\left(1+\sum_{k=1}^{q-1}P_i^k\right)=1+\sum_{l=2}^m\sum_{\alpha_l=1}^{m\choose l}
\sum_{\beta_l=1}^{\nu_l}\Xi_{\alpha_l}^{\beta_l}\,,
\label{expan2}
\ee
where $\Xi_{\alpha_l}^{\beta_l}$ is a product for each period of $l$ Potts spins in configuration $\alpha_l$ with an exponent distribution $\beta_l$. 

The partition function in~\eref{ZNp1} splits in two parts:
\be
\CZ_{N=pm}^{(p)}=\underbrace{\Tr_{\{\tau\}}\e^{-\beta\CH_N^{(p)}[\{\tau\}]}}_{A}+\sum_{l=2}^m\sum_{\alpha_l=1}^{m\choose l}
\sum_{\beta_l=1}^{\nu_l}\underbrace{\Tr_{\{\tau\}}\e^{-\beta\CH_N^{(p)}[\{\tau\}]}\Xi_{\alpha_l}^{\beta_l}}_{B}\,.
\label{ZNp2}
\ee
In $A$, according to~\eref{Trtau}, each of the $pm$ factors in~\eref{embH1} contributes to the trace by:
\be
\frac{\e^{-K}}{q}\Tr_{\tau_j}\left[\e^{qK}+q-1+\left(\e^{qK}-1\right)
\sum_{k=1}^{q-1}\tau_j^k\right]=\e^{-K}\left(\e^{qK}+q-1\right)\,.
\label{TrA}
\ee
In $B$, for each period, $\Xi_{\alpha_l}^{\beta_l}$ contains $l$ supplementary spin terms of the form $\tau_j^{k'}$ 
with $k'=1,2,\ldots,q-1$. Thus the trace involves $p(m-l)$ factors given by~\eref{TrA} and $pl$ factors of the form
\be
\frac{\e^{-K}}{q}\Tr_{\tau_j}\left[\left(\e^{qK}\!+q-1\right)\tau_j^{k'}\!+\left(\e^{qK}\!-1\right)
\sum_{k=1}^{q-1}\tau_j^{k+k'}\right]\!=\e^{-K}\left(\e^{qK}\!-1\right),
\label{TrB}
\ee
where the non-vanishing contribution comes from the term $q-k'$ in the sum over $k$ according to~\eref{Trtau}.
Collecting the different contributions to the partition function, we finally obtain
\bea
&\CZ_{N=pm}^{(p)}=\sum_{l=0}^m{m\choose l}
\nu_l\left[\e^{(q-1)K}\!+\!(q\!-\!1)\,\e^{-K}\right]^{p(m-l)}\!\left[\e^{(q-1)K}\!\!-\e^{-K}\right]^{pl}\nonumber\\
&\ \ \ \ \ \ \ \ \ =\!\left[\e^{(q-1)K}\!+\!(q\!-\!1)\,\e^{-K}\right]^N\left[1\!+\!\sum_{l=2}^m{m\choose l}
\nu_l\left(\frac{\e^{qK}-1}{\e^{qK}\!+\!q\!-\!1}\right)^{pl}\right],
\label{ZNp3}
\eea
where $\nu_l$, given by~\eref{nul}, is such that $\nu_0=1$ and $\nu_1=0$. For $q=2$ 
\be
\nu_l=\left\{
\begin{array}{ll}
1 & \mbox{when $l$ is even}\\
\ms
0 & \mbox{when $l$ is odd}
\end{array}
\right.
\label{nulq2}
\ee
and the Ising result, equation (3.13) in \cite{turban16}, is recovered.

Let $\tens{T}$ be the transfer matrix from $|\sigma_j\sigma_{j+1}\ldots\sigma_{j+m-2}\rangle$ to 
$|\sigma_{j+1}\sigma_{j+2}\ldots\sigma_{j+m-1}\rangle$. As discussed in appendix C, its $m$th power is real and symmetric. 
The real eigenvalues of $\tens{T}^m$, $\omega_l$, and their degeneracy, $g_l$, can be deduced from 
the expression~\eref{ZNp3} of the partition function (see~\eref{eigm}).

\section{Zero-field correlation function}
In this section the zero-field correlation function is obtained for free BC and $m>1$.
The correlations between the Potts variables at sites $i$ and $i'$ are evaluated by taking 
the thermal average of the following expression:
\be
\frac{q\delta_q(s_i-s_{i'})-1}{(q-1)}\,.
\label{sisi'}
\ee
It  is equal to one when the two sites are in the same state and has a vanishing average 
in a fully disordered system. Making use of~\eref{delq} and~\eref{sigmaj} the numerator 
in~\eref{sisi'} can be expressed in terms of Potts spins as:
\be
q\delta_q(s_i-s_{i'})-1=\sum_{k=1}^{q-1}\exp\left[\frac{2i\pi k(s_i-s_{i'})}{q}\right]
=\sum_{k=1}^{q-1}\sigma_i^k\sigma_{i'}^{*k}\,.
\label{qdm1}
\ee
Let us first suppose that $i'=i+rm$. Taking into account~\eref{tt} one may write
\be
\sigma_i\sigma_{i+rm}^*=\prod_{r'=0}^{r-1}\tau_{r'm+i}\tau_{r'm+i+1}^*\,,
\label{sisiprm}
\ee
and the correlation function takes the following form:
\be\fl
\CG_N^{(f)}(i,i\!+\!rm)=\left\langle\frac{q\delta_q(s_i-s_{i+rm})-1}{(q-1)}\right\rangle
=\Tr_{\{\tau\}}\frac{\e^{-\beta\CH_N^{(f)}[\{\tau\}]}}{(q\!-\!1)\CZ_N^{(f)}}
\sum_{k=1}^{q-1}\prod_{r'=0}^{r-1}\tau_{r'm+i}^k\tau_{r'm+i+1}^{*k}\,.
\label{g1}
\ee
Following the same steps that led to~\eref{embH1}, the Boltzmann factor for free BC can be written as
\bea
\fl\e^{-\beta\CH_N^{(f)}[\{\tau\}]}&=\left[\frac{\e^{(q-1)K}+(q-1)\e^{-K}}{q}\right]^{N-m+1}\,
\prod_{j=1}^{N-m+1}\left[1+\frac{\e^{qK}-1}{\e^{qK}+q-1}
\sum_{k'=1}^{q-1}\tau_{j}^{k'}\right]\nonumber\\
\fl&=\frac{\CZ_N^{(f)}}{q^N}\prod_{j=1}^{N-m+1}
\left[1+\frac{\e^{qK}-1}{\e^{qK}+q-1}
\sum_{k'=1}^{q-1}\tau_{j}^{k'}\right]
\label{embH2}
\eea
where the expression of $\CZ_N^{(f)}$ in~\eref{ZNf} has been used
\footnote[3]{Taking the trace over the $N$ Potts spins in \eref{embH2}, all the terms in the product 
involving $\tau_j$ vanish and the trace over 1 gives $q^N$.}.
Inserting this expression in~\eref{g1}, one obtains: 
\bea
\CG_N^{(f)}(i,i+rm)&=\frac{1}{q^N(q-1)}\sum_{k=1}^{q-1}\Tr_{\{\tau\}}
\prod_{r'=0}^{r-1}\tau_{r'm+i}^k\tau_{r'm+i+1}^{*k}\times\nonumber\\
\fl&\ \ \ \ \ \ \ \ \ \times\prod_{j=1}^{N-m+1}\left[1+\frac{\e^{qK}-1}{\e^{qK}+q-1}
\sum_{k'=1}^{q-1}\tau_{j}^{k'}\right],
\label{g2}
\eea
The trace over $\{\tau\}$ contains $r$ factors with $j=r'm+i$ of the form
\be\fl
\Tr_{\tau_{r'm+i}}\left[\tau_{r'm+i}^k+\frac{\e^{qK}-1}{\e^{qK}+q-1}
\sum_{k'=1}^{q-1}\tau_{r'm+i}^{k+k'}\right]=q\,\frac{\e^{qK}-1}{\e^{qK}+q-1}\,,
\label{Trtau1}
\ee
the only non-vanishing contribution coming from the second term for $k'=q-k$ according to~\eref{Trtau}.
The same result is obtained for the $r$ factors with $j=r'm+i+1$ and $k'=k$. 
The trace over the remaining $N-2r$ Potts spins contributes a factor $q^{N-2r}$, 
the sum over $k$ gives $q-1$, so that, finally:
\be
\CG_N^{(f)}(i,i+rm)=\left[\frac{\e^{qK}-1}{\e^{qK}+q-1}\right]^{2r}=\exp\left(-\frac{rm}{\xi}\right)\,.
\label{g3}
\ee
As expected, this expression can be rewritten in terms of transfer matrix eigenvalues~\eref{eigm}
as $(\omega_2/\omega_0)^r$.
The correlation length, given by
\be 
\xi=\frac{m}{2}\left[\ln\left(\frac{\e^{qK}+q-1}{\e^{qK}-1}\right)\right]^{-1}\,,
\label{xi}
\ee
diverges at the zero-temperature critical point when $K\to\infty$.

Let us now consider the case where $i'-i$ is not a multiple of $m$. Using the Potts spin 
variables~\eref{sigmaj} and~\eref{tauj} the inverse transformation in~\eref{sj} translates into:
\be
\sigma_i=\prod_{r'=0}^p\tau_{r'm+i}\tau_{r'm+i+1}^*\,,\qquad i+l=N-pm\,,\qquad l=0,\ldots,m-1\,.
\label{sigmaj1}
\ee
In the same way let
\be
\sigma_{i'}^*=\!\prod_{r'=0}^{p-r}\tau_{r'm+i'}^*\tau_{r'm+i'+1}\,,\quad i'+l'
=N\!-\!pm+rm\,,\quad l'=0,\ldots,m\!-\!1\,,
\label{sigmaj'}
\ee
with, in both cases, $\tau_j=1$ when $j>N$. Since $i'-i=rm+l-l'$, we need $l\neq l'$. In the product 
$\sigma_i\sigma_{i'}^*$, the last factor contributed by $\sigma_i$ is either $\tau_{N-l+1}^*$ or $\tau_N$ 
when $l=0$ whereas for $\sigma_{j'}^*$ it is either $\tau_{N-l'+1}$ or $\tau_N^*$ when $l'=0$. 
Thus these factors cannot all disappear in the product when $l\neq l'$. At least one of them leads to
a vanishing trace over $\{\tau\}$ in the correlation function since the product over $j$ in the Boltzmann 
factor~\eref{embH2} ends at $N-m+1$. It follows that:
\be
\CG_N^{(f)}(i,i')=0\,,\qquad i'-i\neq rm\,.
\label{g4}
\ee
When $m=2$ and $q=2$ this argument no longer applies. With $m=2$ the $\tau_j$ and the $\tau_j^*$ always 
appear twice in the product $\sigma_i\sigma_{i'}^*$ for values of $j\ge i'$. Accordingly, 
the correlation function does not vanish since $\tau_j^2=\tau_j^{*2}=1$ when $q=2$. The 
difference between $q=2$ and $q>2$ when $m=2$ can be understood by looking at the behaviour of 
the correlations in the ground state. For $q=2$ there are 2 degenerate ground states which, 
using Potts variables, are given by $00000\ldots$ and $11111\ldots$ so that 
$\langle2\delta_q(s_i-s_{i'})-1\rangle=1$. When $q=3$, for example, there are 3 degenerate 
ground states, $00000\ldots$, $12121\ldots$ and $21212\ldots$, leading to 
$\langle3\delta_q(s_i-s_{i'})-1\rangle=0$ when $i'-i$ is odd.

\section{Self-duality under external field}
In this section standard methods~\cite{savit80,turban82a,deng10} are used to show that the Potts 
chain with multi-site interactions and periodic BC is self-dual under external field.

According to~\eref{Hs}, the partition function is given by:
\be\fl
\CZ_N^{(p)}(K,H)=\e^{-N(K+H)}\Tr_{\{s\}}\prod_{j=1}^N\exp\left[qK\delta_q
\left(\sum_{l=0}^{m-1}s_{j+l}\right)\right]\exp\left[qH\delta_q(s_j)\right]\,.
\label{ZNKH}
\ee
Introducing the auxiliary function
\be
C(X,x)=\e^{qX}-1+q\delta_q(x)\,,
\label{C}
\ee
one obtains the identity:
\be
\e^{qX\delta_q(y)}=1+\left(\e^{qX}-1\right)\delta_q(y)
=\frac{1}{q}\sum_{x=0}^{q-1}C(X,x)\exp\left(\frac{2i\pi xy}{q}\right)\,.
\label{id}
\ee
Thus the partition function can be rewritten as:
\bea
\CZ_N^{(p)}(K,H)&=\frac{\e^{-N(K+H)}}{q^N}\Tr_{\{s\}}\prod_{j=1}^N
\sum_{u_j=0}^{q-1}\sum_{v_j=0}^{q-1}C(K,u_j)C(H,v_j)\nonumber\\
&\ \ \ \ \ \ \ \ \ \ \ \ \ \ \ \ \ \ \times\frac{1}{q}
\exp\left[\frac{2i\pi}{q}\left(v_js_j+u_j\sum_{l=0}^{m-1}s_{j+l}\right)\right]\,.
\label{ZNKH1}
\eea
Regrouping the factors containing $s_i$ in the last exponential and reordering the sums, one obtains
\bea
\fl\CZ_N^{(p)}(K,H)&=\frac{\e^{-N(K+H)}}{q^N}\Tr_{\{u,v\}}\prod_{j=1}^NC(K,u_j)C(H,v_j)
\prod_{i=1}^N\frac{1}{q}\sum_{s_i=0}^{q-1}\exp\left(\frac{2i\pi s_iw_i}{q}\right)\nonumber\\
\fl&=\frac{\e^{-N(K+H)}}{q^N}\Tr_{\{u,v\}}\prod_{j=1}^NC(K,u_j)C(H,v_j)
\prod_{i=1}^N\delta_q(w_i)\,,
\label{ZNKH2}
\eea
where $w_i$ stands for $v_i+\sum_{l=0}^{m-1}u_{i-l}$. 

\begin{figure}[t!]
\begin{center}
\includegraphics[width=10.522cm,angle=0]{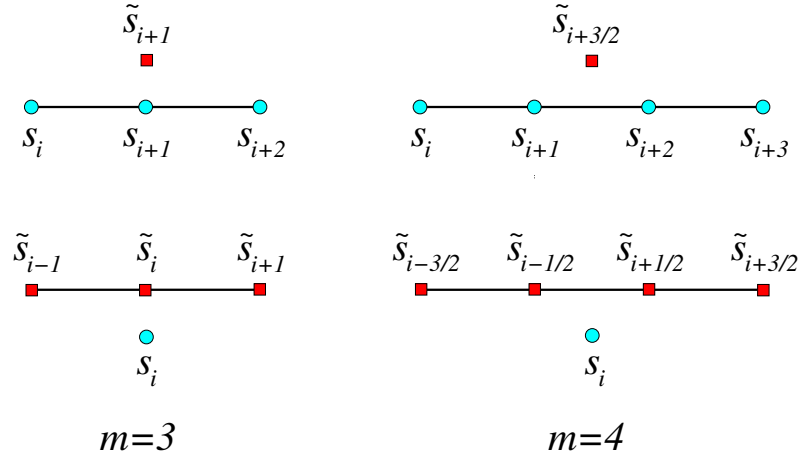}
\end{center}
\vglue -.3cm
\caption{Position of the dual Potts variables (squares) entering in the definitions~\eref{uvts} 
of $u_i$ and $v_i$ relative to the original ones (circles) for odd and even values of $m$. 
\label{fig2}  
}
\end{figure}

Non-vanishing contributions to the partition function correspond to configurations 
of $\{u\}$ and $\{v\}$ such that $w_i=0\!\!\!\pmod{q}\ \forall i$. 
Introducing dual $q$-state Potts variables $\{\ts\}$, 
this condition is automatically satisfied when $u_i$ and $v_i$ take the following forms
\be
u_i=-\ts_{i+(m-1)/2}\pmod{q}\,,\qquad v_i=\sum_{l=0}^{m-1}\ts_{i+l-(m-1)/2}\pmod{q}\,,
\label{uvts}
\ee
such that:
\bea
w_i&=\sum_{l=0}^{m-1}\ts_{i+l-(m-1)/2}-\sum_{l=0}^{m-1}\ts_{i-l+(m-1)/2}\pmod{q}\nonumber\\
&=\sum_{l=0}^{m-1}\ts_{i+l-(m-1)/2}-\sum_{l'=0}^{m-1}\ts_{i-(m-1-l')+(m-1)/2}=0\pmod{q}\,.
\label{om}
\eea
The dual lattice coincides with the original lattice when $m$ is odd. It is shifted 
by half a lattice spacing when $m$ is even (see figure~\ref{fig2}).

Introducing the dual Potts variables in~\eref{ZNKH2}, one obtains:
\be\fl
\CZ_N^{(p)}(K,H)=\frac{\e^{-N(K+H)}}{q^N}\Tr_{\{\ts\}}\prod_{j=1}^NC(K,\ts_{j+(m-1)/2})
C\left(H,\sum_{l=0}^{m-1}\ts_{j+l-(m-1)/2}\right)\,.
\label{ZNKH3}
\ee
Let us rewrite the auxiliary function $C$ as:
\be\fl
C(X,x)=D(X)\exp\left[\tY(q\delta_q(x)-1)\right]
=D(X)\,\e^{-\tY}\left[1+(\e^{q\tY}-1)\delta_q(x)\right]\,.
\label{C1}
\ee
A comparison with~\eref{C} leads to
\be
D(X)=\e^{\tY}\left(\e^{qX}-1\right)\,,\qquad (\e^{qX}-1)(\e^{q\tY}-1)=q\,.
\label{D}
\ee
Making use of these relations, with $\tY=\tH$ when $X=K$ and $\tY=\tK$ when $X=H$, 
the following duality relations for the couplings are obtained:
\be
(\e^{qK}-1)(\e^{q\tH}-1)=q\,,\qquad(\e^{qH}-1)(\e^{q\tK}-1)=q\,.
\label{dua}
\ee
The partition function~\eref{ZNKH3} is now given by:
\be
\CZ_N^{(p)}(K,H)\!=\frac{\e^{-N(K+H)}\e^{N(\tK+\tH)}}{q^N}\left(\e^{qK}\!-\!1\right)^N\!
\left(\e^{qH}\!-\!1\right)^N\!\CZ_N^{(p)}(\tK,\tH)\,.
\label{ZNdua}
\ee
Using~\eref{dua}, this can be put in the more symmetric form:
\be\fl
\frac{\e^{N(K+H)}}{[(\e^{qK}-1)(\e^{qH}-1)]^{N/2}}\CZ_N^{(p)}(K,H)
=\frac{\e^{N(\tK+\tH)}}{[(\e^{q\tK}-1)
(\e^{q\tH}-1)]^{N/2}}\CZ_N^{(p)}(\tK,\tH)\,.
\label{ZNdua1}
\ee
Taking the product of the duality relations in~\eref{dua} and separating the original and dual parts gives
\be
\frac{(\e^{qK}-1)(\e^{qH}-1)}{q}
=\frac{q}{(\e^{q\tK}-1)(\e^{q\tH}-1)}\,,
\label{dua1}
\ee
so that the line $\left(\e^{qK}-1\right)\left(\e^{qH}-1\right)=q$ in the $(K,H)$-plane, 
which is invariant in the duality transformation, is a 
self-duality line.

\section{Mapping on a 2D \boldmath{$q$}-state Potts model when \boldmath{$H\neq0$}}

Let us consider a Potts chain with $N$ spins, $m>1$ and free BC. According to~\eref{Hsigma}, 
the Hamiltonian of the system in an external field $H$ is given by:
\be
-\beta\CH_N[\{\sigma\}]=K\sum_{j=1}^{N-m+1}\sum_{k=1}^{q-1}\prod_{l=0}^{m-1}\sigma_{j+l}^k
+H\sum_{j=1}^N\sum_{k=1}^{q-1}\sigma_j^k\,.
\label{Hsigma1}
\ee
\begin{figure}[t!]
\begin{center}
\includegraphics[width=10.261cm,angle=0]{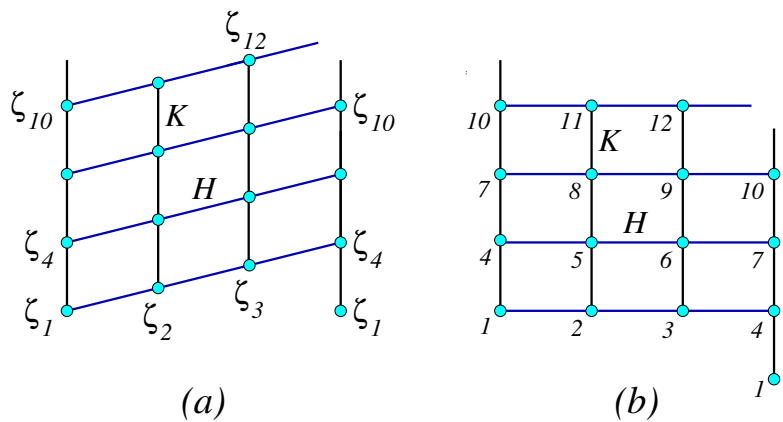}
\end{center}
\vglue -.3cm
\caption{(a) Under the change of spin variables~\eref{zeta} the 1D Potts model 
with $m$-spin interactions in an external field is mapped onto a 2D standard Potts
model on a cylinder with helical BC. $H$ and $K$ are the first-neighbour interactions 
along the helix and parallel to its axis, respectively. The helix has a length $N/m$ 
and $m$ spins per turn. The helicity factor is $1/m$. (b) The same lattice in a 
rectangular representation.   
}
\label{fig3ab}  
\end{figure}
Let us define new Potts spins $\{\zeta\}$ and Potts variables $\{z\}$ such that:
\be
\zeta_j=\exp\left(\frac{2i\pi z_j}{q}\right)=\prod_{i=j}^{N}\sigma_i\,,\qquad z_j=0,\ldots,q-1\,.
\label{zeta}
\ee
Using~\eref{tauj} one obtains
\be\fl
\sigma_j=\left\{
\begin{array}{cc}
\zeta_j\zeta_{j+1}^*\,, & j<N\\
\ms
\zeta_j\,, & j=N
\end{array}
\right.
\,,\qquad
\prod_{l=0}^{m-1}\sigma_{j+l}=\left\{
\begin{array}{cc}
\zeta_j\zeta_{j+m}^*\,, & j<N-m+1\\
\ms
\zeta_j\,, & j=N-m+1
\end{array}
\right.
\,,
\label{zetainv}
\ee
and the correspondence with the original variables is one-to-one. 
The Hamiltonian~\eref{Hsigma1} now takes the following form:
\be\fl
-\beta\CH_N[\{\zeta\}]\!\!=\!\!K\!\sum_{j=1}^{N-m}\sum_{k=1}^{q-1}\!\zeta_j^k{\zeta^*}^k_{\!\!\!j+m}
\!+\!H\sum_{j=1}^{N-1}\sum_{k=1}^{q-1}\!\zeta_j^k{\zeta^*}^k_{\!\!\!j+1}
\!+\!K\sum_{k=1}^{q-1}\!\zeta_{N-m+1}^k\!+\!H\sum_{k=1}^{q-1}\!\zeta_N^k.
\label{Hzeta1}
\ee
Alternatively, using
\be\fl 
\sum_{k=1}^{q-1}\zeta_j^k{\zeta^*}^k_{j'}=\sum_{k=0}^{q-1}
\exp\left[\frac{2i\pi k}{q}\left(z_j-z_{j'}\right)\right]-1
=q\delta_{z_j,z_{j'}}-1\,,\qquad\sum_{k=1}^{q-1}\zeta_j^k=q\delta_{z_j,0}-1\,,
\label{zetaz}
\ee
the following standard form is recovered:
\bea
-\beta\CH_N[\{z\}]&=K\sum_{j=1}^{N-m}\left(q\delta_{z_j,z_{j+m}}-1\right)
+H\sum_{j=1}^{N-1}\left(q\delta_{z_j,z_{j+1}}-1\right)\nonumber\\
&\ \ \ \ \ \ \ \ \ \ \ \ \ \ \ \ \ \ +K\left(q\delta_{z_{N-m+1},0}-1
\right)+H\left(q\delta_{z_N,0}-1\right)\,.
\label{Hz}
\eea
Thus the 1D Potts model with $m$-site interaction $K$ in a field $H$ 
is mapped onto an anisotropic 2D Potts 
model, with standard first-neighbour interactions, on a cylinder with helical BC (see figure~\ref{fig3ab}). 
The interaction is $K$ parallel to the cylinder axis and $H$ along the helix. Local fields $K$ 
and $H$ are acting on two of the end spins. The length of the system is ${\ell}=N/m$, there are $m$ spins per 
turn and the helicity factor is equal to $1/m$.

In the limit ${\ell}=N/m\to\infty$ the free energy of the 1D Potts chain with multi-site 
interactions $K$ under external field $H$ develops a 2D Potts critical singularity along the self-duality
line, $(\e^{qK}-1)(\e^{qH}-1)=q$, when $m\to\infty$~\footnote[4]{The external fields acting on end 
spins do not affect the bulk behaviour.}. Exact expressions for the bulk free energy per site 
have been obtained for the 2D Potts model on its critical line~\cite{baxter73b,baxter78,baxter82}. 
Taking into account the difference in the form of the interactions, the critical free energy 
per site is given by~\cite{baxter82}
\be\fl 
\beta f_b(K,H)=\lim_{m\to\infty}\lim_{N/m\to\infty}-\frac{\ln\cal{Z}_N}{N}=K+H+\psi\,,\quad
\psi=-\frac{1}{2}\ln q-\phi(x_K)-\phi(x_H)\,,
\label{fbKH}
\ee
where
\be
x_K=q^{-1/2}\left(\e^{qK}-1\right)\,,\qquad x_H=q^{-1/2}\left(\e^{qH}-1\right)\,,
\label{x1x2}
\ee
and $x_Kx_H=1$ for the critical system. The transition is second-order when $q\le4$ and first-order
when $q>4$~\cite{baxter73b}. The expression of the function $\phi(x)$ in the different regimes 
can be found in~\cite{baxter82}.
 
Note that successive derivatives of the free energy with respect to $H$, leading to the 
magnetization and the susceptibility for the Potts chain, give the contributions of one type of bonds 
to the internal energy and the specific heat of the 2D Potts model. The derivatives with respect to $K$ 
are of the same nature for both systems. It follows that along the critical line, in the thermodynamic limit 
($\ell\to\infty,m\to\infty$), the thermal and magnetic critical behaviours of the 1D Potts model 
with multi-site interactions in a field, are both governed by the thermal sector of 2D Potts model. 
When $q>4$ the discontinuities of the magnetization and the internal energy 
add up to give the latent heat of the 2D system. When $q\leq4$ the thermal and magnetic critical exponents 
of the second-order phase transition are the 2D thermal Potts exponents~\cite{dennijs79,black81,nienhuis82}.

According to~\eref{zetainv} the two-spin correlation function of the original 1D system
\be
{\cal G}^{(f)}_N(i,i')=\frac{1}{q-1}\sum_{k=1}^{q-1}\langle\sigma_i^k\sigma_{i'}^{*k}\rangle\,,
\label{g5}
\ee
becomes a four-spin correlation function in 2D:
\be
{\cal G}^{(f)}_N(i,i')=\frac{1}{q-1}\sum_{k=1}^{q-1}\langle\zeta_i^k\zeta_{i+1}^{*k}
\zeta_{i'}^{*k}\zeta_{i'+1}^{k}\rangle\,,\qquad i<i'<N\,.
\label{g6}
\ee

When $H=0$, the 2D lattice breaks into $m$ independent spin chains and when $i'=i+rm$ a four-spin
average becomes a product of two-spin averages on two neigbouring chains (see figure~\ref{fig3ab}):
\be
\langle\zeta_i^k\zeta_{i+1}^{*k}\zeta_{i+rm}^{*k}\zeta_{i+rm+1}^{k}\rangle=
\langle\zeta_i^k\zeta_{i+rm}^{*k}\rangle\langle\zeta_{i+1}^k\zeta_{i+rm+1}^{*k}\rangle^*\,.
\label{g7}
\ee
Actually these averages do not depend on $k$ and each factor corresponds to the correlation function 
for two spins at a distance $r$ on a Potts chain with standard first-neighbour interactions
\be
\langle\zeta_i\zeta_{i+rm}^*\rangle=\left(\frac{\e^{qK}-1}{\e^{qK}+q-1}\right)^r\,,
\label{g8}
\ee
from which~\eref{g3} is recovered. When $i'\neq i+rm$, provided $m$ and $q$ are not both equal to two,
the four-spin average in~\eref{g6} always involve some vanishing factor.

\section{Other multi-site Potts models}

\begin{figure}[t!]
\begin{center}
\includegraphics[width=12.304cm,angle=0]{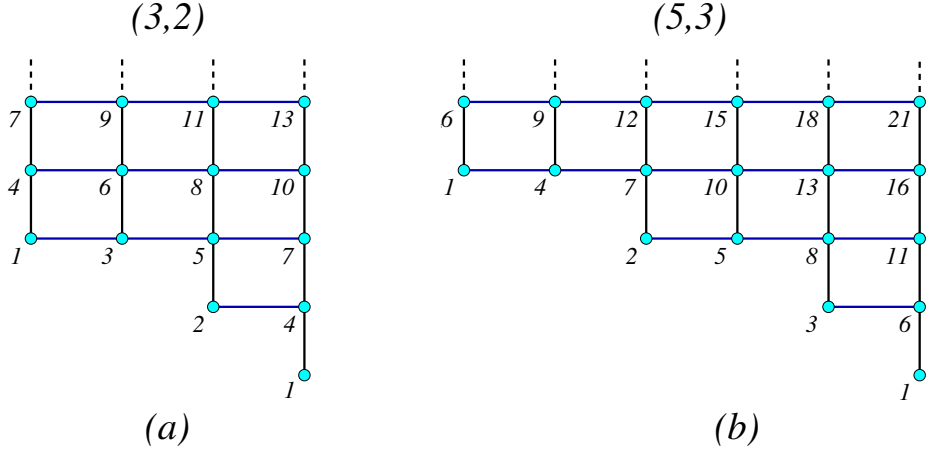}
\end{center}
\vglue -.3cm
\caption{Examples in the rectangular representation of 2D lattices associated with the transformed 
Hamiltonian~\eref{Hzeta2} when $m$ and $n$ are mutually primes. The helicity factor is~$n/m$.
\label{fig4ab}  
}
\end{figure}

We consider now a 1D Potts model with free BC and two types of multi-site 
interactions~\footnote[5]{This type of Hamiltonian is also self-dual as shown 
more generally in~\cite{deng10} for a simple hypercubic lattice.}. 
In this (m,n) Hamiltonian, with $m>n>1$, the external field term is replaced by a $n$-site interaction:
\be
-\beta\CH_N[\{\sigma\}]=K\sum_{j=1}^{N-m+1}\sum_{k=1}^{q-1}\prod_{l=0}^{m-1}\sigma_{j+l}^k
+L\sum_{j=1}^{N-n+1}\sum_{k=1}^{q-1}\prod_{l=0}^{n-1}\sigma_{j+l}^k\,.
\label{Hsigma2}
\ee
The change of variables~\eref{zeta} leads to the following transformed Hamiltonian: 
\be\fl
-\beta\CH_N[\{\zeta\}]\!\!=\!\!K\!\sum_{j=1}^{N-m}\sum_{k=1}^{q-1}\!\zeta_j^k{\zeta^*}^k_{\!\!\!j+m}
\!+\!L\sum_{j=1}^{N-n}\sum_{k=1}^{q-1}\!\zeta_j^k{\zeta^*}^k_{\!\!\!j+n}
\!+\!K\sum_{k=1}^{q-1}\!\zeta_{N-m+1}^k\!+\!L\sum_{k=1}^{q-1}\!\zeta_{N-n+1}^k.
\label{Hzeta2}
\ee

\begin{figure}[t!]
\begin{center}
\includegraphics[width=13.957cm,angle=0]{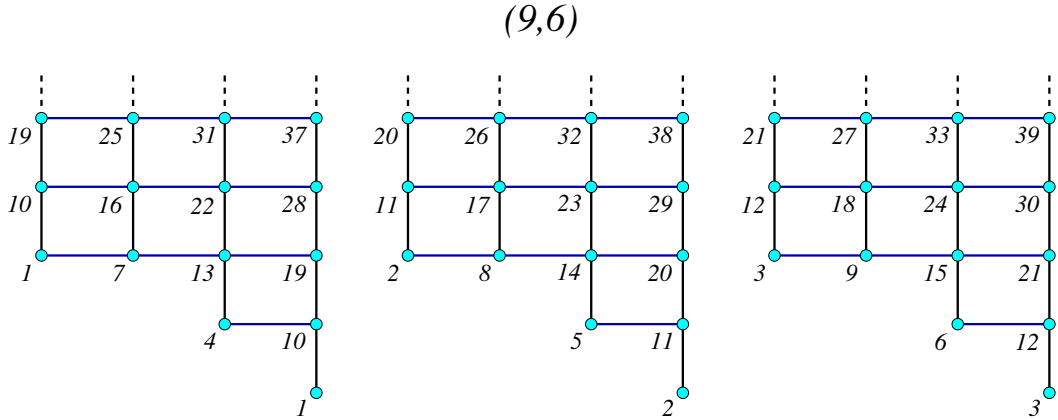}
\end{center}
\vglue -.3cm
\caption{When $m=fm'$ and $n=fn'$, with $m'$ and $n'$ mutually primes, the Hamiltonian~\eref{Hzeta2} splits into $f$ 
non-interacting parts to which correspond $f$ independent 2D lattices with 
helical BC and $m'$ spins per turn. The expressions of the lattice length, $\ell=N/m$, and the helicity 
factor, $n/m=n'/m'$, remain unchanged.
\label{fig5}  
}
\end{figure}

For the $(m,n)$ Hamiltonian, in the rectangular lattice representation (figure~\ref{fig4ab}), 
the horizontal interaction $L$ couples spins $\zeta_j$ and $\zeta_{j+n}$ thus generating $n$ chains 
of connected sites with $j=0,1,\ldots,n-1$ (mod $n$). When $m$ and $n$ are mutually 
primes these chains are connected by vertical interactions between spins $\zeta_j$
and $\zeta_{j+m}$. Starting from site $j$ one reaches site $j+mn$ via either $m$ horizontal steps or $n$ 
vertical steps. Thus the 2D lattice has helical BC, $m$ steps per turn and the helicity factor is $n/m$.

Let us now consider the case where $m$ and $n$ have a greatest common factor $f$ so that $m=fm'$, $n=fn'$, 
with $m'$ and $n'$ mutually primes (figure~\ref{fig5}). Then among the $n$ horizontal chains 
of connected spins with $j=0,1,\ldots,n-1$ (mod $n$) the $f$ chains with $j=0,1,\ldots,f-1$ (mod $f$) 
belong to $f$ distinct 2D lattices since with $m=0$ (mod $f$) there are no vertical interconnections. 
Starting from site $j$ one can reach site $j+fn'm'$ through either $m'$ horizontal steps or $n'$ 
vertical steps on the same lattice. The $f$ distinct 2D lattices, with length $N/m$, have helical BC, $m'$ 
steps per turn and their helicity factor remains equal to $n/m=n'/m'$.

Note that Potts chains with more complex multi-site interactions can be mapped onto triangular 
or honeycomb lattices as shown in appendix~D.

\section{Conclusion}

In this work we have used some spin transformation to obtain exact results for the 
zero-field partition functions and the two-spin correlation function of a $q$-state Potts chain with multi-site
interactions. We have shown that the model is self-dual under external field. With another spin 
transformation, the Potts chain with $m$-site interaction $K$ in a field $H$ has been mapped onto a standard
2D $q$-state Potts model with first-neighbour interactions $K$ and $H$. The 2D system with $N$ spins has a length 
$\ell=N/m$, a transverse size $m$ and helical BC in the transverse direction. 

Thus the Potts chain in a field develops a critical singularity on the self-duality line, 
$(\e^{qK}-1)(\e^{qH}-1)=q$, as $\ell\to\infty$ and $m\to\infty$, i.e., in the thermodynamic 
limit for the 2D system. Along this line the thermal and magnetic critical behaviours 
of the Potts chain are both governed by the thermal critical behaviour of the 2D Potts model. 
The transition is first-order when $q>4$ and second-order when $q\leq4$. 

A numerical exploration of the finite-size scaling behaviour
on the self-duality line would be of interest. The development of the critical singularities 
with increasing values of $N$ and $m$ should be studied for some fixed values of the aspect 
ratio~$\ell/m=N/m^2$.


\appendix

\section{Clock angular variables}
\setcounter{section}{1}
\label{inter}

Using \eref{delq} the Potts multi-site interaction in~\eref{Hs} can be rewritten as 
\bea
\fl q\delta_q\left(\sum_{l=0}^{m-1}s_{j+l}\right)-1&=\sum_{k=1}^{q-1}\exp\left(\frac{2i\pi k}{q}
\sum_{l=0}^{m-1}s_{j+l}\right)=\frac{1}{2}\sum_{k=1}^{q-1}\left[\exp\left(\frac{2i\pi k}{q}
\sum_{l=0}^{m-1}s_{j+l}\right)+\mbox{c.c.}\right]\nonumber\\
\fl&=\sum_{k=1}^{q-1}\cos\left(\frac{2\pi k}{q}
\sum_{l=0}^{m-1}s_{j+l}\right)
\eea
or, introducing the clock angular variable $\theta_j=2\pi s_j/q=0,2\pi/q,\ldots,2\pi(q-1)/q$,
\be
q\delta_q\left(\sum_{l=0}^{m-1}s_{j+l}\right)-1
=\sum_{k=1}^{q-1}\cos\left(k\sum_{l=0}^{m-1}\theta_{j+l}\right)\,.
\label{int2}
\ee
Similarly for the field term $q\delta_q\left(s_j\right)-1=\sum_{k=1}^{q-1}\cos\left(k\theta_j\right)$.

\section{Calculation of \boldmath{$\nu_l$}}
\label{anul}

Let us consider a term in the expansion~\eref{expan} with $l$ spins per period: 
\be
\prod_{r=0}^{p-1}\prod_{i=1}^{l}\tau_{rm+j_i}^{k_i}\,.
\label{term}
\ee
For the number of distinct distributions of the exponents $k_i=1,\ldots,q-1$, 
such that $\sum_{i=1}^{l}k_i=0\!\pmod{q}$, we find:
\bea
\fl\nu_l&=\!\!\!\!\sum_{k_1,k_2,\ldots,k_l=1}^{q-1}\!\!\!\delta_q\left(\sum_{i=1}^lk_i\right)
=\frac{1}{q}\sum_{k=0}^{q-1}\prod_{i=1}^l
\underbrace{\sum_{k_i=1}^{q-1}\exp\left(\frac{2i\pi kk_i}{q}\right)}_{q\delta_q(k)-1}
=\frac{1}{q}\sum_{k=0}^{q-1}\left[\,q\delta_q(k)-1\right]^l\nonumber\\
\fl&=\frac{1}{q}\left[(q-1)^l+(-1)^l(q-1)\right]\,.
\label{nul1}
\eea
Thus $\nu_0=1$ and $\nu_1=0$, independent of $q$.
For $q=2$, due to the fact that $k_i=1$ for Ising spins, one obtains $\nu_{2k+1}=0$ and $\nu_{2k}=1$. 

Note that the value of $\nu_l$ in~\eref{nul1} leads to a total number of terms in~\eref{expan} given by
\be\fl
1+\sum_{l=2}^m{m\choose l}\nu_l
=\frac{1}{q}\sum_{l=0}^m{m\choose l}\left[(q-1)^l+(-1)^l(q-1)\right]
=\frac{1}{q}\sum_{l=0}^m{m\choose l}(q-1)^l=q^{m-1}
\label{total}
\ee
as required. 

As an illustration let us look for the form of the expansion when $m=3$ and $q=4$. With $x,y,z$ 
standing for the product of $p$ spins, $\prod_{r=0}^{p-1}\tau_{rm+j}$, with respectively, $j=1,2,3$, 
so that $x^q=y^q=z^q=1$, we obtain:
\bea
\fl(1+xy^3+x^2y^2+x^3y)(1+yz^3+y^2z^2+y^3z)=1\nonumber\\
+(xy^3+x^2y^2+x^3y)+(yz^3+y^2z^2+y^3z)+(zx^3+z^2x^2+z^3x)\nonumber\\
+(xyz^2+yzx^2+zxy^2+x^3y^3z^2+y^3z^3x^2+z^3x^3y^2)\,.
\label{expan1}
\eea
On the right-hand side the terms in brackets correspond to the different exponent 
distributions for the same spin configuration. The values $\nu_2=3$ and $\nu_3=6$ are 
in agreement with~\eref{nul1}.

\section{Transfer matrix at \boldmath{$H=0$}}
\label{trans}

Before considering general values of $q$ and $m$, let us study the properties of 
the transfer matrix of a 3-state Potts model with 3-site interactions at $H=0$.

In the basis
$\{|00\rangle,|01\rangle,|02\rangle,|10\rangle,|11\rangle,|12\rangle,|20\rangle,|21\rangle,|22\rangle\}$,
the transfer matrix of the Hamiltonian~\eref{Hsp} from $|s_js_{j+1}\rangle$ to $|s_{j+1}s_{j+2}\rangle$, 
takes the following form:
\be
\tens{T}=\left(
\begin{array}{ccccccccc}
\e^{2K} & \e^{-K} & \e^{-K} &  0      &  0      &  0      &  0      &  0      &  0      \\
 0      &  0      &  0      & \e^{-K} & \e^{-K} & \e^{2K} &  0      &  0      &  0      \\
 0      &  0      &  0      &  0      &  0      &  0      & \e^{-K} & \e^{2K} & \e^{-K} \\
\e^{-K} & \e^{-K} & \e^{2K} &  0      &  0      &  0      &  0      &  0      &  0      \\
 0      &  0      &  0      & \e^{-K} & \e^{2K} & \e^{-K} &  0      &  0      &  0      \\
 0      &  0      &  0      &  0      &  0      &  0      & \e^{2K} & \e^{-K} & \e^{-K} \\  
\e^{-K} & \e^{2K} & \e^{-K} &  0      &  0      &  0      &  0      &  0      &  0      \\ 
 0      &  0      &  0      & \e^{2K} & \e^{-K} & \e^{-K} &  0      &  0      &  0      \\
 0      &  0      &  0      &  0      &  0      &  0      & \e^{-K} & \e^{-K} & \e^{2K} 
 \end{array}
\right)\,.
\label{T}
\ee 
It is asymmetric and has complex eigenvalues:
\bea
\fl\lambda_0&=\e^{2K}+2\e^{-K}\,,\qquad \lambda_{2,k}
=\left[(\e^{2K}+2\e^{-K})(\e^{2K}-\e^{-K})^2\right]^{1/3}\e^{2ik\pi/3}
\,,\quad k=0,1,2\,,\nonumber\\
\fl\lambda_3&=\e^{2K}-\e^{-K}\,.
\label{eig}
\eea
Both $\lambda_{2,k}$ and $\lambda_3$ are doubly degenerate. The oscillating behaviour is linked 
to the periodicity of the degenerate ground states. With $a=\e^{6K}+2\e^{-3K}$, $b=\e^{3K}+\e^{-3K}+1$,
the cube of $\tens{T}$, corresponding to a transfer by one period from $s_j$ to $s_{j+3}$, 
leads to the symmetric matrix 
\be
\tens{T}^3=\left(
\begin{array}{ccccccccc}
 a      &  b      &  b      &  b      &  3      &  b      &  b      &  b      &  3      \\
 b      &  a      &  b      &  b      &  b      &  3      &  3      &  b      &  b      \\
 b      &  b      &  a      &  3      &  b      &  b      &  b      &  3      &  b      \\
 b      &  b      &  3      &  a      &  b      &  b      &  b      &  3      &  b      \\
 3      &  b      &  b      &  b      &  a      &  b      &  b      &  b      &  3      \\  
 b      &  3      &  b      &  b      &  b      &  a      &  3      &  b      &  b      \\ 
 b      &  3      &  b      &  b      &  b      &  3      &  a      &  b      &  b      \\
 b      &  b      &  3      &  3      &  b      &  b      &  b      &  a      &  b      \\
 3      &  b      &  b      &  b      &  3      &  b      &  b      &  b      &  a     
 \end{array}
\right)\,,
\label{T3}
\ee 
with real eigenvalues
\be\fl
\omega_0=(\e^{2K}+2\e^{-K})^3\,,\quad \omega_2=(\e^{2K}+2\e^{-K})(\e^{2K}-\e^{-K})^2
\,,\quad\omega_3=(\e^{2K}-\e^{-K})^3\,.
\label{eig3}
\ee
$\omega_0$ is non degenerate and the two last eigenvalues are, respectively, six times 
and two times degenerate.

For any value of $m$ and $q$, the $q^{m-1}$ eigenvalues of $\tens{T}^m$, $\omega_l$, 
and their degeneracy, $g_l$, can be extracted from 
the expression of the partition function with periodic BC. Since
\be
\CZ_{N=mp}^{(p)}=\Tr(\tens{T}^{m})^p=\sum_lg_l\,\omega_l^p
\label{ZNp4}
\ee
it follows from~\eref{ZNp3} that
\be\fl
\omega_l=\left[\e^{(q-1)K}\!+(q-1)\e^{-K}\right]^m\left[\frac{\e^{qK}\!-1}{\e^{qK}\!+q-1}\right]^l\,,
\quad g_l=\!{m\choose l}\nu_l\,,\quad l=0,2,\ldots,m\,,
\label{eigm}
\ee
with $\nu_l$ given by~\eref{nul1}.

\section{Triangular and honeycomb lattices}
\label{latt}

\begin{figure}[t!]
\begin{center}
\includegraphics[width=14.391cm,angle=0]{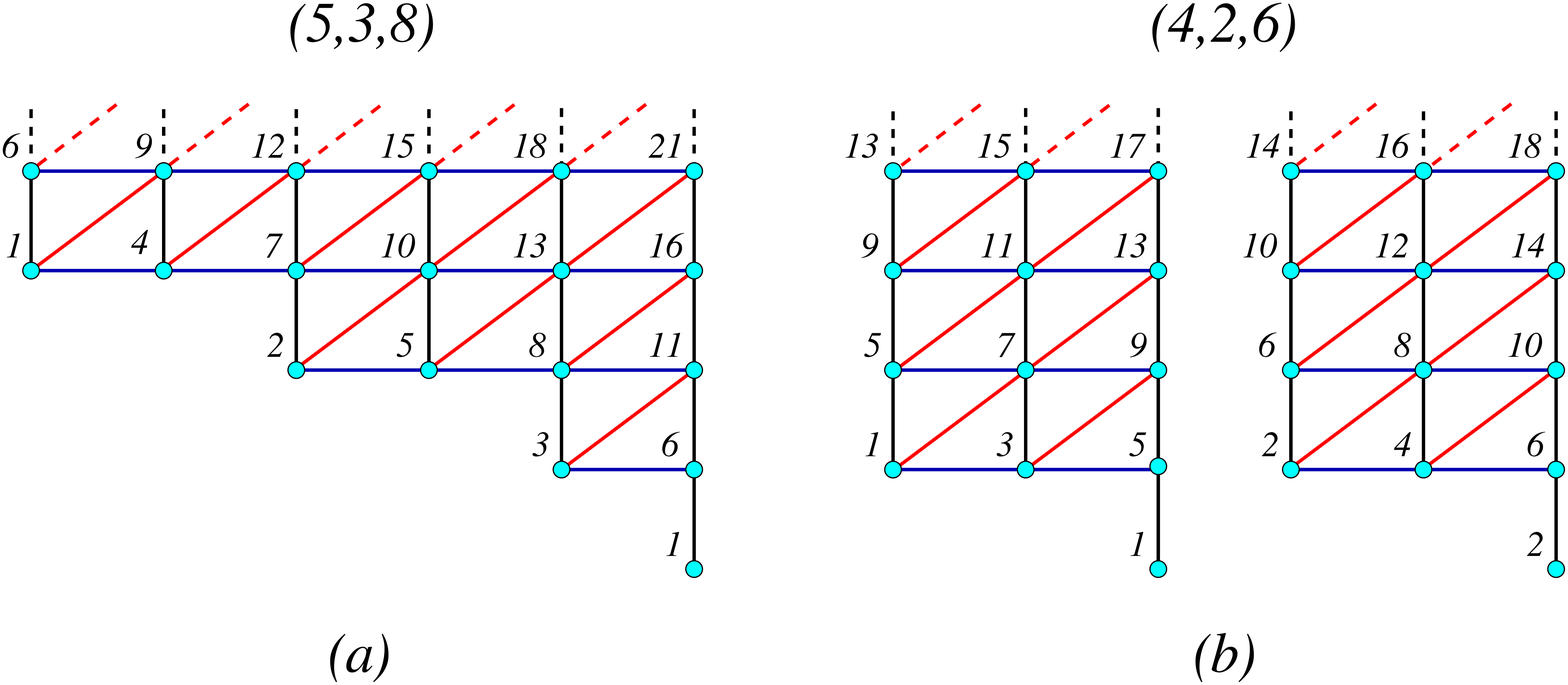}
\end{center}
\vglue -.3cm
\caption{Rectangular representation of 2D lattices associated with 
the transformed Hamiltonian~\eref{Hzeta3} when (a) 
$m$ and $n$ are mutually primes and (b) $m$ and $n$ have a greatest common factor $f$. 
The first-neighbour interactions are $K$, $L$ and $M$ in the vertical, 
horizontal and diagonal directions, respectively.
\label{fig6ab}  
}
\end{figure}

\begin{figure}[t!]
\begin{center}
\includegraphics[width=6.7391cm,angle=0]{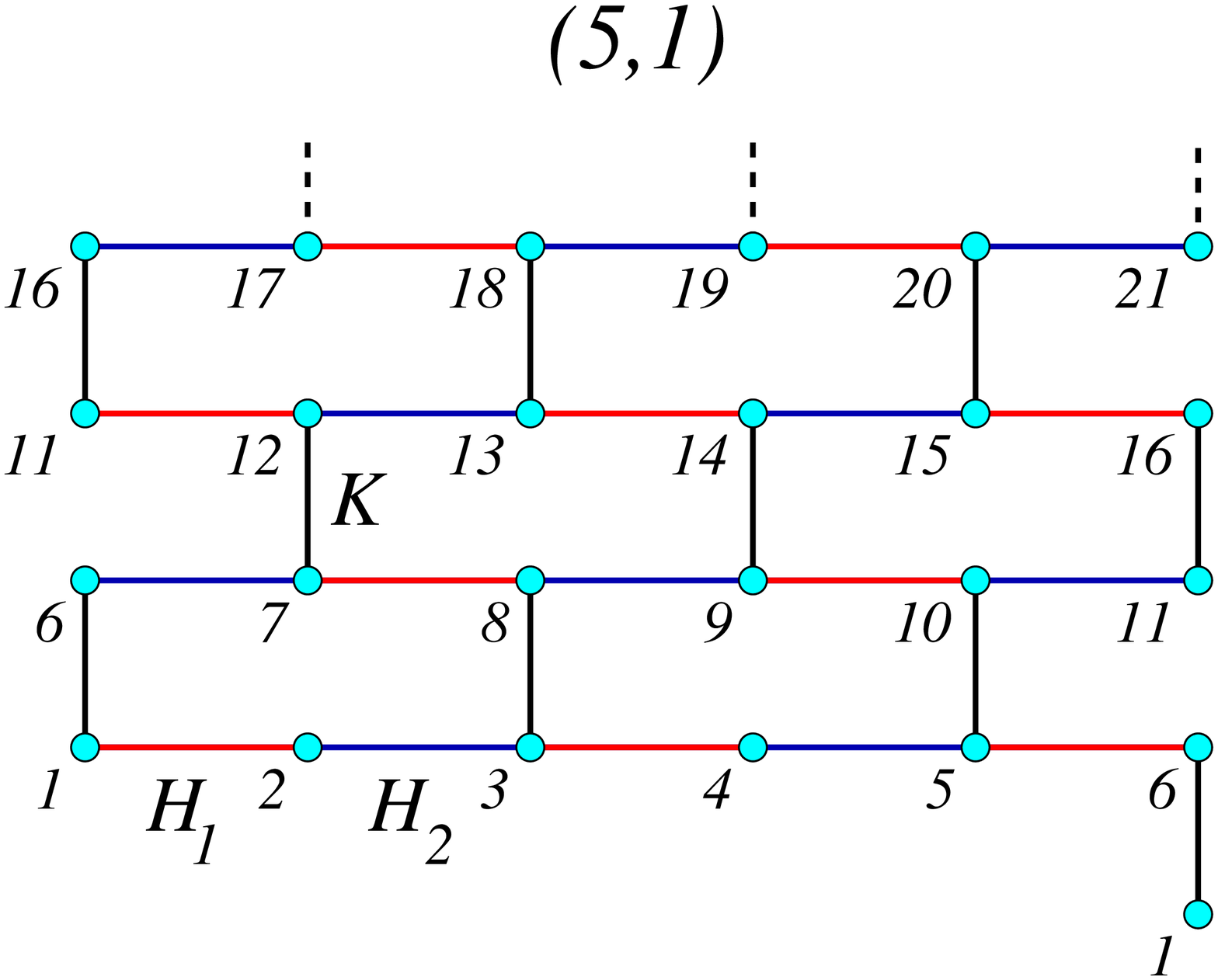}
\end{center}
\vglue -.3cm
\caption{Rectangular representation of the honeycomb lattice associated 
with the transformed Hamiltonian~\eref{Hzeta4}.
\label{fig7}  
}
\end{figure}

With the $(m,n,m+n)$ Hamiltonian ($m>n$) such that
\be\fl
-\beta\CH_N[\{\sigma\}]=K\!\!\!\sum_{j=1}^{N-m+1}\sum_{k=1}^{q-1}\prod_{l=0}^{m-1}\!\sigma_{j+l}^k
+L\!\!\!\sum_{j=1}^{N-n+1}\sum_{k=1}^{q-1}\prod_{l=0}^{n-1}\!\sigma_{j+l}^k
+M\!\!\!\!\!\!\sum_{j=1}^{N-m-n+1}\sum_{k=1}^{q-1}\prod_{l=0}^{m+n-1}
\!\!\!\sigma_{j+l}^k\,,
\label{Hsigma3}
\ee
the change of variables~\eref{zeta} leads to the following transformed Hamiltonian: 
\bea
\!\!\!\!-\beta\CH_N[\{\zeta\}]&=K\!\sum_{j=1}^{N-m}\sum_{k=1}^{q-1}
\!\zeta_j^k{\zeta^*}^k_{\!\!\!j+m}
+L\!\sum_{j=1}^{N-n}\sum_{k=1}^{q-1}\zeta_j^k{\zeta^*}^k_{\!\!\!j+n}+M\!\!\!\!
\sum_{j=1}^{N-m-n}\sum_{k=1}^{q-1}\zeta_j^k{\zeta^*}^k_{\!\!\!j+m+n}\nonumber\\
&\ \ \ \ \ \ \ \ \ \ \ +\!K\sum_{k=1}^{q-1}\!
\zeta_{N-m+1}^k\!+\!L\sum_{k=1}^{q-1}\!\zeta_{N-n+1}^k\!
+\!M\sum_{k=1}^{q-1}\!\zeta_{N-m-n+1}^k.
\label{Hzeta3}
\eea
As shown in figure~\ref{fig6ab}-a when $m$ and $n$ are mutually primes it corresponds 
to a triangular lattice Potts model with first-neighbour interactions
on a cylinder with helical BC, an external fields acting on 
three end spins. When $m$ and $n$ have a greatest common factor $f$, 
as in figure \ref{fig6ab}-b, $f$ independent triangular lattices are obtained.
 
Finally let us consider a 1D Potts model with $m$-spin interaction $K$ ($m>1$) starting 
on odd sites only, and two external fields, $H_1$ and $H_2$, acting on 
odd and even sites. When $N-m$ is even the Hamiltonian can be written as:
\be\fl
-\beta\CH_N[\{\sigma\}]=K\!\!\!\!\!\!\!\sum_{p=1}^{(N-m)/2+1}\sum_{k=1}^{q-1}
\prod_{l=0}^{m-1}\!\sigma_{2p+l-1}^k
+H_1\!\!\!\sum_{p=1}^{\lfloor (N+1)/2\rfloor}\sum_{k=1}^{q-1}\sigma_{2p-1}^k
+H_2\!\sum_{p=1}^{\lfloor N/2\rfloor}\sum_{k=1}^{q-1}\sigma_{2p}^k\,.
\label{Hsigma4}
\ee
The transformed Hamiltonian then takes the following form
\bea
\fl&-\beta\CH_N[\{\zeta\}]=K\!\sum_{p=1}^{(N-m)/2}
\sum_{k=1}^{q-1}\!\zeta_{2p-1}^k{\zeta^*}^k_{\!\!\!2p+m-1}+\!K\sum_{k=1}^{q-1}\!\zeta_{N-m+1}^k
+H_1\sum_{p=1}^{\lfloor N/2\rfloor}\sum_{k=1}^{q-1}
\zeta_{2p-1}^k{\zeta^*}^k_{\!\!\!2p}\nonumber\\
\fl&\ \ \ \ \ \ \ \ \ \ \ \ \ \ \ \ \ \ \ \ \ \ \ \ \ \ +\left\{
\begin{array}{ll}
H_2\sum_{p=1}^{N/2-1}\sum_{k=1}^{q-1}\zeta_{2p}^k{\zeta^*}^k_{\!\!\!2p+1}
+H_2\sum_{k=1}^{q-1}\!\zeta_N^k\,, \quad& \mbox{$N$ even}\\
\ms\ms
H_1\sum_{k=1}^{q-1}\!\zeta_N^k+ H_2\sum_{p=1}^{\lfloor N/2\rfloor}
\sum_{k=1}^{q-1}\zeta_{2p}^k{\zeta^*}^k_{\!\!\!2p+1}\,,&\mbox{$N$ odd}
\end{array}
\right.\,,
\label{Hzeta4}
\eea
which corresponds to a Potts model with first-neighbour 
interactions on the honeycomb lattice as shown in figure~\ref{fig7}.

\end{document}